\documentclass[12pt,a4paper,twoside,groupcitations]{article}
\usepackage[T1]{fontenc}
\usepackage[ansinew]{inputenc}
\usepackage[english]{babel}
\usepackage{amsfonts,bm}
\usepackage{amsmath}
\usepackage{array}
\usepackage{amsthm}
\usepackage{amssymb}
\usepackage{graphicx}
\usepackage{braket}
\usepackage{eucal}
\usepackage{verbatim}
\usepackage[table]{xcolor}
\usepackage{caption}
\usepackage{cite}
\usepackage{textcomp}
\usepackage{mathrsfs}
\usepackage{bbold}
\usepackage{array,booktabs}
\usepackage{siunitx}
\usepackage{longtable}
\usepackage{multicol}
\DeclareMathOperator{\Tr}{Tr}
\DeclareMathOperator{\tr}{tr}
\newcommand{\dd }{{\mathrm d}}
\title{\bf Quantum fields in teleparallel gravity: renormalization at one-loop}
\author{Roberto~Casadio$^{ab}$\thanks{casadio@bo.infn.it},
$\ $
Iber\^e Kuntz$^{ab}$\thanks{kuntz@bo.infn.it}
$\ $
and
Gregorio~Paci$^{a}$\thanks{gregorio.paci@studio.unibo.it}
\\
\\
$^a${\em Dipartimento di Fisica e Astronomia, Universit\`a di Bologna}
\\
{\em via Irnerio~46, 40126 Bologna, Italy}
\\
\\
$^b${\em I.N.F.N., Sezione di Bologna, I.S.~FLAG}
\\
{\em viale B.~Pichat~6/2, 40127 Bologna, Italy}
}
\begin{document}
\date{}
\maketitle
\begin{abstract}
\noindent
We consider the quantization of matter fields in a background described by the teleparallel equivalent to general relativity.
The presence of local Lorentz and gauge symmetries gives rise to different coupling prescriptions, which we analyse
separately.
As expected, quantum matter fields produce divergences that cannot be absorbed by terms
in the background action of teleparallel equivalent to general relativity.
Nonetheless, the formulation of teleparallel gravity allows one to find out the source of the problem.
By imposing local Lorentz invariance after quantization, we show that a modified teleparallel gravity, in which
the coefficients in the action are replaced by free parameters, can be renormalized at one-loop order without
introducing higher-order terms.
This precludes the appearance of ghosts in the theory.
\end{abstract}
%
%
\section{Introduction}
\setcounter{equation}{0}
\label{sec:intro}
General relativity is one of the cornerstones of modern physics.
It successfully describes the classical gravitational interaction in terms of the spacetime curvature,
which results from the presence of matter.
Particles and fields are then forced to evolve in a fully dynamical, and generally curved,
background that is itself a solution of the Einstein field equations.
This state of affairs shows how non-linear the interaction of gravity with matter turns out to be.
\par
The non-linearity of general relativity is indeed the major obstacle to quantizing the gravitational interaction.
Present techniques of quantum field theory, including the construction of the Fock space, largely rely
on linear field equations for the free theory and on the use of perturbation theory otherwise.
Some success had been achieved by quantizing linear perturbations of the metric as originally proposed
by Feynman~\cite{Feynman:1963ax} and DeWitt~\cite{DeWitt:1967ub}, but such an approach was soon
discovered to lead to a non-renormalizable field theory.
It was only with the culmination of effective field theories in recent years~\cite{Donoghue:1994dn}
that quantum general relativity started to be taken seriously.
The effective field theory approach is based on an energy expansion and the dynamical
degrees of freedom are then defined by the lowest order Lagragian.
The Fock space for gravity is therefore defined by the Einstein-Hilbert term,~\footnote{Since the Einstein-Hilbert
Lagrangian is itself non-linear, the construction of the Fock space remains formal and the application of the
effective action is practically limited to classical background geometries representing the ``vacuum'' in the Fock space.
\label{f:1}}
but interactions are non-linear due to higher-order curvature invariants.
This implies that the effective field theory is applicable to gravity only at energies much smaller than the Planck scale,
beyond which the formalism breaks down, thus missing out on all of the fundamental
aspects of quantum gravity that take place in the deep ultraviolet (UV).
\par
The problem of quantizing general relativity could perhaps be ameliorated by using variables
and field strengths other than the metric and the curvature, in an attempt of bringing the gravitational
interaction closer in form to other quantum field theories.
One interesting example of this possibility is teleparallel gravity~\cite{Hayashi:1967se,Pellegrini:1963,Cho:1975dh}
(see Refs.~\cite{Arcos:2004tzt,AldrovandiBook,Bahamonde:2021gfp} for extensive reviews).
In this theory, the fundamental variable is the tetrad field (or the gauge field for translations) and its strength
is measured by the spacetime torsion rather than the curvature.
In spite of the presence of geometrical concepts involved in the definition of spacetime, the gravitational
interaction itself is not geometrized as it is in general relativity.
In this scenario, gravity is a force rather than a manifestation of a curved geometry.
Another important aspect of teleparallel gravity is the separation of the gravitational interaction (tetrad)
from inertia (spin connection), which are all mixed up in the metric of general relativity.
While there is no clear meaning associated to ``quantum inertia'', the procedure for quantizing
the gravitational force can closely follow that adopted in gauge theories.
In fact, as we shall see, teleparallel gravity is nothing else than a complicated gauge theory for the
translation group~\cite{Lucas:2008gs,Pereira:2019woq}.
Taking for granted (some form of) the equivalence principle, another way to understand
the difference between general relativity and teleparallel gravity is that in general relativity one only
considers the metric associated with actual particle motions (described as geodesics of the metric
solving the Einstein field equations), whereas in teleparallel gravity one takes as reference the (Minkowski)
metric which would describe free-particle trajectories if gravity could be switched off.
This latter metric, despite being clearly unphysical (gravity cannot be switched off), can be naturally
associated with the ``absolute vacuum'' of quantum field theory, hence there is hope that standard
quantum field theory techniques applied to teleparallel gravity could lead to progress in quantum gravity.
\par
Despite all the differences with respect to general relativity, the dynamical equations of teleparallel
gravity are equivalent to the Einstein field equations for a particular choice of the Lagrangian
[see Eq.~\eqref{TEGR}].
This equivalence actually extends to the level of the action, where the boundary term in teleparallel gravity
exactly reproduces the Gibbons-Hawking-York term~\cite{Oshita:2017nhn}.
Such an equivalence, nonetheless, has only been shown to hold classically.
Since teleparallel gravity adopts the tetrad (or a gauge potential) as opposed to the metric as its
fundamental field, it is not clear whether the aforementioned equivalence would extend to quantum
scales because (for instance) the path integral is performed with respect to a different variable.
We recall that non-stationary paths in configuration space also contribute to the path integral,
which suggests the violation of this equivalence.
Even in a semiclassical approximation, where gravity is not quantized, quantum matter fluctuations
could induce different interactions with the background, thus departing from the usual result of quantum
fields in a general relativistic geometry.
\par
The purpose of this paper is to study teleparallel gravity in the quantum regime.
We shall adopt a semiclassical approach by considering quantum fields in a classical teleparallel
geometry and calculate the one-loop divergences.
The quantization of matter fields in spacetime with torsion has a long history (see \cite{Buchbinder:1992rb,Shapiro:2001rz} for in-depth reviews). However, we should stress that, unlike other theories with torsion, teleparallel gravity can be formulated as a gauge theory for the group of translations in the affine Minkowski space.
The feature of translational gauge invariance is indeed what makes teleparallel gravity stand out from the other torsional theories. Particularly, the findings of Section~\ref{gaugeonly} cannot be obtained by the simple application of the results of \cite{Buchbinder:1992rb,Shapiro:2001rz} to the teleparallel connection.

Teleparallel gravity is thus invariant under both the
translational gauge group and local Lorentz transformations~\cite{Krssak:2018ywd}.
We shall consider coupling prescriptions with respect to both and we shall show how their one-loop
structure differ.
Our main finding is that the local Lorentz symmetry and the lack of additional free parameters
are the culprits for the non-renormalizability of quantum fields coupled to gravity.
A slight modification in the action of the gravitational sector can however be performed in order
to absorb all one-loop divergences from matter fields if one also imposes local Lorentz symmetry
after quantization (rather than before as usual).
We stress that such a modification does not require higher derivatives, thus no ghosts or instabilities
show up in the theory.
\subsection{Notation}
 In the following we will have to introduce different kinds of connections and covariant derivatives, 
which may be a source of confusion. 
We will mostly employ the notation commonly used to formulate teleparallel theories of gravity, 
and list the main symbols in Table~\ref{tab:notation} for the reader's convenience. 
As a general rule, we will denote objects used in teleparallel gravity with $\bullet$, 
those employed in general relativity with $\circ$, 
whereas we will not use any distinctive symbol for those objects defined in terms of a \emph{generic} connection.
\begin{table}[htb]
    \def\arraystretch{1.5}
    \caption{Notation adopted in this paper} 
    \label{tab:notation}
    \centering 
    \begin{tabular}{>{$}c<{$} c S[table-format=3.5] S[table-format=1.6]} 
        \toprule
        \text{Symbol} & Meaning    \\
        \midrule
        \omega^a_{\  b\mu}    & General spin connection      \\
        \Gamma^{\rho}_{\mu\nu}    & General spacetime connection     \\
        R\,\!^{a}_{\ {b}\nu\mu}    & Riemann curvature tensor of  $\omega^a_{\  b\mu} $              \\
        T\,\!^a_{\mu\nu} & Torsion tensor of   $\omega^a_{\  b\mu} $\\
        \overset{\bullet}{\omega}\,\!^a_{\  e\nu}  & Teleparallel spin connection \\
        \overset{\bullet}{R}\,\!^a_{\ b\mu\nu}  & Riemann curvature tensor of $\overset{\bullet}{\omega}\,\!^a_{\  e\nu}$ \\
        \overset{\bullet}{T}\,\!^a_{\mu\nu} & Torsion tensor of $\overset{\bullet}{\omega}\,\!^a_{\  e\nu}$ \\
        \overset{\circ}{\omega}\,\!^a_{\ b \mu}  & Spin connection of general relativity \\
        \overset{\circ}{\Gamma}\,\!^{\rho}_{\mu\nu}    & General Relativity connection (Levi-Civita) \\
        \overset{\circ}{R}\,\!^{\rho}_{\lambda\nu\mu}    & Riemann curvature tensor of  $\overset{\circ}{\Gamma}\,\!^{\rho}_{\mu\nu}$ \\
        B^a_{\ \mu} & Translational gauge potential \\
        \overset{\circ}{\nabla}_\mu    & Covariant derivative associated with $\overset{\circ}{\Gamma}\,\!^{\rho}_{\mu\nu}$ \\
        \overset{\bullet}{\nabla}_{\mu} & Covariant derivative associated with $ \overset{\bullet}{\omega}\,\!^{ab}_{\ \ \mu} - \overset{\bullet}{K}\,\!^{ab}_{\ \ \mu}=\overset{\circ}{\omega}\,\!^{ab}_{\ \ \mu}$ \\
        \nabla_\mu & Covariant derivative associated with $\overset{\circ}{\Gamma}\,\!^{\rho}_{\mu\nu}$ plus a gauge connection \\
        \Box & d'Alembert operator constructed using $\nabla_\mu$ \\
        h_{\mu} & Gauge covariant derivative associated with $B^a_{\ \mu}$ \\
        \bottomrule
    \end{tabular}
\end{table}
\section{Teleparallel gravity}
\setcounter{equation}{0}
\label{TG}
In order to describe events and processes
occurring in the spacetime, one must specify not only dynamical equations but also the geometrical setting
where the dynamics takes place.
Hence, in this section, we start by reviewing briefly the geometrical setting of teleparallel
gravity and its equivalence with classical general relativity.
\subsection{Geometrical setting}
Mathematically speaking, there are different ways of endowing a set of (spacetime) points with
geometrical structures.
Typical examples of geometrical structures that often appear in physics include the metric tensor
and the connection (that is, the abstract notion of parallelism).
These structures are generally independent from each other, which allows one to layer them up into
a rich environment of geometrical objects.
\par
Generally speaking, the connection can be characterized by three quantities:
curvature, torsion and non-metricity (when spacetime is also endowed with a metric tensor).
The kinematics of general relativity assumes that both torsion and non-metricity vanish,
thus spacetime is solely described in terms of the curvature of the connection.
In this case, and in this case only, the connection is fully determined by the metric,
which allows one to loosely speak about the curvature of the metric.
The connection in this special case is said to be of the Levi-Civita type.
On the other hand, teleparallel gravity employs a geometry where the curvature and
the non-metricity vanish, but torsion is in general not zero.
We shall completely disregard non-metricity in this paper as it plays no role in teleparallel
gravity.~\footnote{For a more general perspective, see {\em e.g.}~\cite{LH} and references
therein.}
\par
More precisely, let us introduce the tetrad fields $h^a = h^a_{\ \mu} \,\dd x^\mu$ by imposing
\begin{equation}
    h^a(h_b) = \delta^a_{\ b}
    \ .
     \label{introducing_tetrads}
\end{equation}
Note that the set $\{h_a\}$ (respectively, $\{h^a\}$) is nothing but a basis for the tangent (respectively,
co-tangent) space, which is obtained by a simple change of basis from the more popular coordinate basis,
namely $\{\partial_\mu\}$ (respectively, $\{\dd x^\mu\}$).
{From the bilinear property of the metric one finds}
\begin{equation}
    \eta_{ab} = h^{\ \mu}_a\, h^{\ \nu}_b\, g_{\mu\nu}
    \ ,
\end{equation}
which is usually employed to introduce the tetrad fields. 
This relation allows one to easily move back and forth from the coordinate to the non-coordinate basis
by contracting expressions with $h^{\ \mu}_a$.
For a given tetrad, one can also introduce the spin connection $\omega^a_{\  b\mu}$, which accounts
for the covariance under Lorentz transformations.
The curvature and torsion associated to $\omega^a_{\  b\mu}$ are respectively defined by
\begin{equation}
    R^a_{\ b\mu\nu}
        =
    \partial_\mu \omega^a_{\  b\nu}
    - \partial_\nu \omega^a_{\  b\mu}
    + \omega^a_{\  e\mu} \,\omega^e_{\  b\nu}
    - \omega^a_{\  e\nu} \,\omega^e_{\  b\mu}
\end{equation}
and
\begin{equation}
    T^a_{\ \mu\nu}
    =
    \partial_\mu h^a_{\ \nu}
    - \partial_\nu h^a_{\ \mu}
    + \omega^a_{\  e\mu} \,h^e_{\ \nu}
    - \omega^a_{\  e\nu} \,h^e_{\ \mu}
    \ .
    \label{torsion}
\end{equation}
The teleparallel (or Weitzenb\"ock) connection $\overset{\bullet}{\omega}\,\!^a_{\  e\nu}$ is such that
$\overset{\bullet}{R}\,\!^a_{\ b\mu\nu} \equiv 0$ and $\overset{\bullet}{T}\,\!^a_{\mu\nu} \neq 0$.
We recall that, following the traditional notation, quantities defined by the teleparallel connection
shall be denoted by a bullet $\bullet$, whereas quantities defined from the Levi-Civita
connection shall be denoted with an open circle $\circ$.
We shall not use any distinctive symbol for objects obtained from a generic connection,
such as the ones appearing in Section~\ref{SD}.
\par
An important result concerns the decomposition of an arbitrary connection as the sum of the Levi-Civita
one and the contortion tensor, to wit
\begin{equation}
    \omega^a_{\ b \mu}
    =
    \overset{\circ}{\omega}\,\!^a_{\ b \mu}
    + K^a_{\ b \mu}
    \ ,
\end{equation}
where $\overset{\circ}{\omega}\,\!^a_{\ b \mu}$ is the usual Levi-Civita connection of general relativity
and
\begin{equation}
    K^{a}_{\ b \mu}
    =
    \frac12
    \left(
        T_{b \ \ \mu}^{\ a} + T_{\mu \ \ b}^{\ a} - T^a_{\ b \mu}
    \right)
    \label{cont}
\end{equation}
denotes the contortion tensor.
One can use the tetrad to transform all Lorentz indices into spacetime ones, in which case the above
relation reads
\begin{equation}
	\Gamma^{\rho}_{\mu\nu}
	=
	\overset{\circ}{\Gamma}\,\!^{\rho}_{\mu\nu} + K^{\rho}_{\mu\nu}
	\label{condec}
	\ .
\end{equation}
The decomposition of the connection induces a similar decomposition on the Riemann tensor,
\begin{equation}
	R^{\rho}_{\ \lambda\nu\mu}
	=
	\overset{\circ}{R}\,\!^{\rho}_{\lambda\nu\mu}
	+ Q^{\rho}_{\ \lambda\nu\mu}
	\ ,
	\label{general curvaure=levi-civita curvature + Q}
\end{equation}
where
\begin{equation}
	\overset{\circ}{R}\,\!^{\rho}_{\lambda\nu\mu}
	=
	\partial_{\nu} \overset{\circ}{\Gamma}\,\!^{\rho}_{\lambda\mu}
	- \partial_{\mu} \overset{\circ}{\Gamma}\,\!^{\rho}_{\lambda\nu}
	+ \overset{\circ}{\Gamma}\,\!^{\rho}_{\gamma\nu} \,\overset{\circ}{\Gamma}\,\!^{\gamma}_{\lambda\mu}
	- \overset{\circ}{\Gamma}\,\!^{\rho}_{\gamma\mu} \,\overset{\circ}{\Gamma}\,\!^{\gamma}_{\lambda\nu}
\end{equation} 
corresponds to the Riemann tensor of the Levi-Civita connection and
\begin{align}
	Q^{\rho}_{\ \lambda\nu\mu}
	=
	&\
	\partial_{\nu} K^{\rho}_{\lambda\mu}
	- \partial_{\mu} K^{\rho}_{\lambda\nu}
	+ \overset{\circ}{\Gamma}\,\!^{\rho}_{\sigma\nu} \,K^{\sigma}_{\lambda\mu}
	- \overset{\circ}{\Gamma}\,\!^{\rho}_{\sigma\mu} \,K^{\sigma}_{\lambda\nu}
	\nonumber
	\\
	&
	- \overset{\circ}{\Gamma}\,\!^{\sigma}_{\lambda\nu} \,K^{\rho}_{\sigma\mu}
	+ \overset{\circ}{\Gamma}\,\!^{\sigma}_{\lambda\mu} \,K^{\rho}_{\sigma\nu}
	+ K^{\rho}_{\sigma\mu} \,K^{\sigma}_{\lambda\nu}
	- K^{\rho}_{\sigma\nu} \,K^{\sigma}_{\lambda\mu}
	\label{Q tensor}
\end{align}
measures the departure from general relativity's curvature due to the presence of torsion. 
Notice that Eq.~\eqref{Q tensor} can be written also in terms of the covariant derivative associated
with the Levi-Civita connection $\overset{\circ}{\nabla}_{\mu}$. 
Moreover, using Eq.~\eqref{condec} for the teleparallel connection, it can be written using only such a connection.
Since the geometrical setting of teleparallel gravity consists of a manifold with
$\overset{\bullet}{R}\,\!^{\rho}_{\lambda\nu\mu}=0$, as corollary of the above we find
\begin{equation}
	\overset{\bullet}{Q}\,\!^{\rho}_{\lambda\nu\mu}
	=
	-\overset{\circ}{R}\,\!^{\rho}_{\lambda\nu\mu}
	\ .
	\label{QR}
\end{equation}
This relation is the stepping stone for the equivalence between general relativity and teleparallel gravity.
Eq.~\eqref{QR} shall also be important for the derivation of the one-loop divergences produced by matter fields
in teleparallel gravity.
\subsection{Gravity as a translational gauge theory}
\label{gauge}
Teleparallel gravity can be viewed as a gauge theory for the group $T_4$
formed by translations in the affine Minkowski space~\cite{Pereira:2019woq}
\begin{equation}
    x^a \to x^a + \epsilon^a(x^\mu)
    \ ,
    \label{shift}
\end{equation}
where $\epsilon^a$ denotes the infinitesimal transformation parameter.
Under a local transformation of the type~\eqref{shift}, a generic field $\varphi$, carrying an arbitrary
representation of the Lorentz group, transforms covariantly as
\begin{equation}
    \delta_\epsilon \varphi
    =
    \epsilon^a(x)\, \partial_a \varphi
    \ ,
\end{equation}
where $\partial_a$ is the generator of translations.
However, derivatives explicitly break the gauge covariance,
\begin{equation}
    \delta_\epsilon(\partial_\mu \varphi)
    =
    \epsilon^a(x)\, \partial_a(\partial_\mu \varphi)
    + \partial_\mu \epsilon^a\, \partial_a \varphi
    \ ,
\end{equation}
thus requiring the introduction of a connection $B^a_{\ \mu}$ in order to retain the invariance under
local translations.
The gauge field then naturally defines a covariant derivative~\footnote{Unlike most teleparallel gravity literature,
our $B^a_{\ \mu}$ is given (Planckian) mass dimension by defining
$(16\,\pi \, G_{\rm N})^{-1/2}\, B^a_{\ \mu} \to B^a_{\ \mu}$, with $G_{\rm N}$ the Newton constant.
This is important to make contact with the standard conventions in quantum field theory.
\label{fn:scale}}
\begin{equation}
    h_\mu \varphi
    =
    \partial_\mu \varphi
    + {\sqrt{16\,\pi \, G_{\rm N} } } B^a_{\ \mu}\, \partial_a \varphi
    \ ,
\end{equation}
{where $G_{\rm N}$ is the Newton constant}, and the minimal coupling prescription
\begin{equation}
    \partial_\mu
    \to
    h_\mu
    =
    h^a_{\ \mu}\, \partial_a
    \ ,
\end{equation}
which turns the underlying theory into a gauge theory under $T_4$ by enforcing the gauge invariance
in all couplings to other sectors.
The gauge field strength, in particular, reads
\begin{equation}
    \left[h_\mu, h_\nu\right]
    =
   \overset{\bullet}{T}\,\!^{a}_{\ \mu\nu}\, \partial_a
    \ ,
\end{equation} 
where
\begin{equation}
     \overset{\bullet}{T}\,\!^{a}_{\ \mu\nu}
    =
    {\sqrt{16\,\pi \, G_{\rm N} } }
    (
    \partial_\mu B^a_{\ \nu}
    - \partial_\nu B^a_{\ \mu}
    )
    \ .
    \label{gst}
\end{equation}
Adding $[\partial_\mu, \partial_\nu] \,x^a = 0$ to Eq.~\eqref{gst}, yields
\begin{equation}
     \overset{\bullet}{T}\,\!^{a}_{\ \mu\nu}
    =
    \partial_\mu h^a_{\ \nu}
    - \partial_\nu h^a_{\ \mu}
    \ ,
\end{equation}
which indeed coincides with the torsion tensor defined in Eq.~\eqref{torsion} for a vanishing spin connection.
The above considerations indeed only hold for the class of proper frames, where the spin connection vanishes
and inertial effects are absent.
General expressions, valid in arbitrary frames, can be easily obtained by introducing a Lorentz
transformation $x^a \to \Lambda^a_{\ b}\,x^b$ in the formulas above.
As a result, the gauge field strength becomes
\begin{equation}
     \overset{\bullet}{T}\,\!^{a}_{\ \mu\nu}
    =
    \partial_\nu h^a_{\ \mu}
    - \partial_\mu h^a_{\ \nu}
    + \overset{\bullet}{\omega}\,\!^a_{\ e\nu} \,h^e_{\ \mu}
    - \overset{\bullet}{\omega}\,\!^a_{\ e \mu} \,h^e_{\ \mu}
    \ ,
\end{equation}
recovering Eq.~\eqref{torsion} for the teleparallel spin connection
and allowing one to interpret the gravitational field strength as torsion.
\par
As in the prototypical Yang-Mills gauge theories, the action for the gravitational sector is given by~\footnote{We adopt the metric signature $(-+++)$.}
\begin{equation}
    S_G
    =
    {-\frac{1}{16\,\pi\,G_{\rm N}} }
    \int
    \tr\left[\overset{\bullet}{T} \wedge \star \overset{\bullet}{T}\right]
    \ ,
    \label{act}
\end{equation}
where the trace $\tr$ acts on the gauge indices and $\wedge$ stands for the wedge product.
The dual operator above, however, is a generalization of the usual Hodge expression, to wit
\begin{equation}
    \star T^\rho_{\ \mu\nu}
    =
    h \,\epsilon_{\mu\nu\alpha\beta}
    \left(
        \frac{1}{4} \,T^{\rho\alpha\beta}
        + \frac{1}{2} \,T^{\alpha\rho\beta}
        - T^{\lambda \alpha}_{\ \lambda} \,g^{\rho\beta}
    \right)
    \ ,
    \label{star}
\end{equation}
where 
\begin{equation}
h = \det( h^a_{\ \mu})
\ ,
\label{deth}
\end{equation}
and the torsion tensor with three spacetime indices is obtained through contraction with the tetrad,
that is $T^\rho_{\ \mu\nu}=h_a^{\ \rho}\ T^a_{\ \mu\nu}$.
This generalized definition of the Hodge dual is required in order to take into account all new contractions
of a tensor that now exist due to the mapping between internal and external indices allowed by the tetrad.
This so-called {\em soldered property\/} of the internal space is indeed the crucial difference between
Yang-Mills theories, in which gauge and spacetime indices are completely unrelated, and teleparallel gravity.
\par
From Eqs.~\eqref{act} and \eqref{star} one finds
\begin{equation}
	\overset{\,\, \bullet}{\mathcal{L}}
	=
	{-\frac{h}{16\,\pi\,G_{\rm N}} }
	\left(
	    \frac{1}{4}\,\overset{\bullet}{T}\,\!^{\rho}_{\ \mu\nu}\, \overset{\bullet}{T}\,\!_{\rho}^{\ \mu\nu}
	    + \frac{1}{2}\,\overset{\bullet}{T}\,\!^{\rho}_{\ \mu\nu} \,\overset{\bullet}{T}\,\!^{\nu\mu}_{\ \ \rho}
	    - \overset{\bullet}{T}\,\!^{\rho}_{\ \mu\rho}\,\overset{\bullet}{T}\,\!^{\nu\mu}_{\ \ \nu}
	\right)
	\ ,
	\label{TEGR}
\end{equation}
where we have taken the Minkowski metric $\eta_{a b}$ for the internal space.
In Eq.~\eqref{TEGR}, the first term corresponds to the usual Yang-Mills term, whereas the others are new contractions
only possible for soldered spaces.
The dynamical equivalence between general relativity and teleparallel gravity can be easily shown from
Eqs.~\eqref{cont} and \eqref{QR}, which yields
\begin{equation}
	\left(
	     \frac{1}{4}\,\overset{\bullet}{T}\,\!^{\rho}_{\ \mu\nu}\, \overset{\bullet}{T}\,\!_{\rho}^{\ \mu\nu}
	    + \frac{1}{2}\,\overset{\bullet}{T}\,\!^{\rho}_{\ \mu\nu} \,\overset{\bullet}{T}\,\!^{\nu\mu}_{\ \ \rho}
	    - \overset{\bullet}{T}\,\!^{\rho}_{\ \mu\rho}\,\overset{\bullet}{T}\,\!^{\nu\mu}_{\ \ \nu}
	\right)
	+ \frac{2}{h}\,\partial_{\mu}\!\left(  h \overset{\bullet}{T}\,\!^{\nu\mu}_{\ \ \nu}  \right)
	=
	- \overset{\circ}{R}
		\ ,   
	\label{eq:TGGReq}
\end{equation}
where $\overset{\circ}{R}$ stands for the Ricci scalar of the Levi-Civita connection.
Thus, the Lagrangian density~\eqref{TEGR} only differs from the Einstein-Hilbert Lagrangian by a boundary term,
which does not affect the equations of motion.
One can also show that this boundary term exactly reproduces the Gibbons-Hawking-York one~\cite{Oshita:2017nhn}.
\par
The coupling prescription also acquires a correction in order to compensate for the change of frames due to the
local Lorentz invariance.
Ultimately, the full coupling prescription due to local translations and local Lorentz transformations
reads~\cite{Krssak:2018ywd}
\begin{equation}
	\partial_{\mu} \Phi
	\to
	\overset{\bullet}{\nabla}_{\mu} \Phi
	=
	\partial_{\mu}\Phi
	- \frac{i}{2}\left( \overset{\bullet}{\omega}\,\!^{ab}_{\ \ \mu}
	- \overset{\bullet}{K}\,\!^{ab}_{\ \ \mu}  \right)
	S_{ab} \, \Phi
	\ ,
	\label{fullc}
\end{equation}
where $S_{ab}$ is the Lorentz generator in the representation of the field $\Phi$.
Because of the relation~\eqref{condec}, it is immediate to see that this coupling is equivalent to the standard
one that follows from the equivalence principle of general relativity. 
\par
The separation of the gravitational degrees of freedom from inertial effects makes it clear that only
$B^a_{\ \mu}$ (and not $\omega^a_{\ \mu\nu}$) should be quantized.
In quantum field theory approaches to general relativity, such a distinction between gravity and inertia
is not manifested, thus one ends up quantizing them both altogether.
The spin connection is however not dynamical and its quantization is thus bounded to frame-dependent
spurious effects. These effects also occur when considering matter fields, as in the present work. 
Indeed, they also become manifest when one adopts the full-coupling prescription,
in which local Lorentz invariance is imposed before the quantization of matter fields.
As we will see, if we invert the order of this procedure by imposing the local Lorentz coupling prescription
only after quantization, all one-loop divergences can be renormalized by the terms already present in the
bare action. 
\par
In any case, we shall see later that the divergent part of the one-loop effective action
is purely geometrical and can therefore be considered as a contribution to the gravitational sector
of the theory rather than to matter.
\section{Effective action and the Schwinger-DeWitt formalism}
\setcounter{equation}{0}
\label{SD}
There are two different ways of computing the one-loop divergences of a quantum field theory:
perturbation theory in the coupling constants (which is the standard textbook approach popularized
by the Feynman diagrams) and the Schwinger-DeWitt technique.
The latter has the advantage of being covariant and becomes especially useful in gravity.
In this section, we shall just recall the main results of this covariant method developed by Schwinger and
DeWitt~\cite{Schwinger:1951nm,DeWitt:1965} (see also Refs.~\cite{Barvinsky:1985an,Avramidi:2000bm}
for in-depth presentations), in which the one-loop effective action is expanded in inverse powers of
a mass parameter.~\footnote{This mass parameter $m$ can be the actual mass of field's quanta
or a constant term, such as a constant curvature or the minimum of some potential.
Its presence is important in order to kill off the proper-time integral for large $s$,
thus allowing for an asymptotic expansion for small $s$ (large $m$).
Without such a mass parameter, the Schwinger-DeWitt expansion becomes inapplicable.}
Equivalent covariant approaches were also developed
in Refs.~\cite{tHooft:1974toh,Gilkey:1975iq,Dowker:1976zf,Hawking:1976ja}.
\par
Using the background field formalism, one can perform a semiclassical expansion of the effective action
\begin{equation}
    \Gamma[\varphi]
    =
    S[\varphi]
    + \hbar \, \Gamma^{(1)}[\varphi]
    + \mathcal O(\hbar^2)
    \ ,
\label{eq:S}
\end{equation}
where $\varphi$ collectively denotes the set of background fields (of any spin), $S[\varphi]$ stands for their classical action,
\begin{equation}
    \Gamma^{(1)}[\varphi]
    =
    \frac{i}{2} \Tr \log F(\nabla)
    \label{TrLog}
\end{equation}
and
\begin{equation}
    F(\nabla)\,\delta(x,y)
    =
    \frac{\delta^2 S}{\delta\varphi(x)\,\delta\varphi(y)}
    \ ,
\end{equation}
where
\begin{equation}
    \Tr f
    =
    \int\dd^dx
    \tr \braket{x|f(x)|x}
\end{equation}
denotes the functional trace and $\tr$ the trace of finite-dimensional operators (running over internal indices).
The particular form of the operator $F(\nabla)$ obviously depends on the classical action, but for the majority of
the cases of physical interest one has the so-called minimal operators~\footnote{We have singled
out the non-minimal coupling and the mass for convenience, thus $\hat P$ denotes the remaining potential.}
\begin{equation}
    F(\nabla)
    =
     \Box
    + \hat P
    - \frac{\overset{\circ}{R}}{6} \,\mathbb 1
    - m^2 \, \mathbb 1
    \ ,
    \label{hessian}
\end{equation}
where $\hat P$ is the potential, $\mathbb 1$ is the identity in the internal space, $m$ is a mass parameter,
$\Box = g^{\mu\nu} \,\nabla_\mu \,\nabla_\nu$ and $\nabla_\mu$ is the covariant derivative with respect
to a general connection that comprises \emph{both} the Levi-Civita and the gauge ones (not to be confused with
$\overset{\circ}{\nabla}_\mu$ which is a \emph{pure} Levi-Civita connection, see Table~\ref{tab:notation}).
The only property of such a connection that is necessary for the present formalism is the corresponding curvature.
Notice that $\nabla_\mu$ is defined with respect to both spacetime and gauge connections,
thus the commutator $[\nabla_\mu, \nabla_\nu]$ contains information about the Riemann tensor and the vector
bundle curvature ({\em i.e.},~the gauge field strength).
In fact, the commutator produces different results when acting on different types of objects.
For an arbitrary spacetime vector $u^\alpha$, we have
\begin{equation}
    \left[\nabla_\mu, \nabla_\mu\right] u^\alpha
    =
    \overset{\circ}{R}\,\!^\alpha_{\ \beta\mu\nu}\,u^\beta
    \ ,
    \label{curv1}
\end{equation}
whereas for a bundle vector $\phi^a$, we find
\begin{equation}
    \left[\nabla_\mu, \nabla_\mu\right] \phi^a
    =
    \mathcal R^a_{\ b\mu\nu}\, \phi^b
    \equiv
    (\hat{\mathcal R}_{\mu\nu})^a_{\ b} \, \phi^b
    \ .
    \label{curv2}
\end{equation}
The application of $\nabla_\mu$ on mixed objects containing both spacetime and internal indices
thus must be taken with respect to both connections.
The expression for the trace-log formula in~\eqref{TrLog} above is only formal as it requires
the use of some regularization scheme to make sense.
We shall adopt the dimensional regularization here.
\par
The main outcome of the Schwinger-DeWitt formalism is an explicit expression for the divergent part of 
the one-loop effective action, $\Gamma^{(1)}_\text{div}$, in terms of only a few coefficients~\cite{DeWitt:1965}, 
which then determine the structure of the necessary counterterms. 
These coefficients can be denoted as $\hat a_n$, where $n$ is an integer number.
As we will see soon, the divergences only take place up to $n=d/2$, where $d$ is the spacetime dimension.
Thus, we shall only need the first three coefficients $\hat a_n$ in four dimensions.
They read
\begin{align}
    \hat a_0(x,x)
    &=
    \mathbb 1
   \label{HAMIDEW0}
    \\
    \hat a_1(x,x)
    &=
    \hat P
    \label{HAMIDEW1}
    \\
    \hat a_2(x,x)
    &=
    \frac{1}{180} \left(\overset{\circ}{R}_{\alpha\beta\mu\nu} \,\overset{\circ}{R}\,\!^{\alpha\beta\mu\nu}
    -\overset{\circ}{R}_{\mu\nu} \,\overset{\circ}{R}\,\!^{\mu\nu} +  \Box \overset{\circ}{R}\right) \mathbb 1
    + \frac{1}{2} \,\hat P^2
    + \frac{1}{12} \,\hat{\mathcal R}_{\mu\nu} \,\hat{\mathcal R}^{\mu\nu}
    + \frac{1}{6}\, \Box \hat P
    \ .
    \label{HAMIDEW2}
\end{align}
Using this result, one finally finds the divergent part of the
effective action
\begin{align}
    \Gamma^{(1)}_\text{div}
    &=
    \frac{1}{16\,\pi^2\,\varepsilon}
    \int\mathrm{d}^4x \,g^{1/2}\,
    \tr [\hat a_2(x,x)]
    \nonumber
    \\
    &=
    \frac{1}{32\,\pi^2\,\varepsilon}
    \int\mathrm{d}^4x\, g^{1/2}\,
    \tr \left[
        \frac{1}{90} \left(
            \overset{\circ}{R}\,\!_{\alpha\beta\mu\nu} \,\overset{\circ}{R}\,\!^{\alpha\beta\mu\nu}
            - \overset{\circ}{R}_{\mu\nu} \,\overset{\circ}{R}\,\!^{\mu\nu}
        \right) \mathbb 1
        + \hat P^2
        + \frac{1}{6} \,\hat{\mathcal R}_{\mu\nu} \,\hat{\mathcal R}^{\mu\nu}
    \right]
    \ ,
    \label{div}
\end{align}
where $\varepsilon = 4 - d$ and we have dropped total derivatives.
Note that $\Gamma^{(1)}_\text{div}$ is dimensionless, therefore $\hbar \,\Gamma^{(1)}_\text{div}$
has the correct dimensions of $\hbar$ for an action.
From Eqs.~\eqref{eq:TGGReq} and \eqref{eq:S}, one then finds the one-loop effective action~\footnote{We recall
that we keep gravity classical in this work, hence it is represented by the Einstein-Hilbert term in Eq.~\eqref{gammaLp}.
Moreover, in light of Eq.~\eqref{eq:TGGReq}, the gravitational sector is fully written in terms of torsion
in the case of teleparallel gravity.}
\begin{eqnarray}
\Gamma
&\!\!=\!\!&
\frac{1}{16\,\pi\,G_{\rm N}} 
\int \mathrm{d}^4 x\,
h\,\overset{\circ}{R}
+ S_m
+
\hbar\,
\Gamma^{(1)}
\nonumber
\\
&\!\!=\!\!&
{-\frac{1}{16\,\pi\,G_{\rm N}} }
\int \mathrm{d}^4 x\,h\,
\left(
	  \frac{1}{4}\,\overset{\bullet}{T}\,\!^{\rho}_{\ \mu\nu} \,\overset{\bullet}{T}\,\!_{\rho}^{\ \mu\nu}
	  + \frac{1}{2}\,\overset{\bullet}{T}\,\!^{\rho}_{\ \mu\nu} \,\overset{\bullet}{T}\,\!^{\nu\mu}_{\ \ \rho}
	  - \overset{\bullet}{T}\,\!^{\rho}_{\ \mu\rho}\,\overset{\bullet}{T}\,\!^{\nu\mu}_{\ \ \nu}
	  \right)
+ S_m 
+
\hbar \,\Gamma^{(1)}
\ ,
\label{gammaLp}
\end{eqnarray}
where we split the gravitational background $S_{G}$ from the matter sector $S_m$ in the classical action 
$S = S_{G} + S_m$ [see Eq.~\eqref{eq:S}].
This shows that the counter-terms required for cancelling out divergences must be suppressed by the Planck
length $\ell_{\rm p}^2 = 16 \, \pi \, G_{\rm N} \, \hbar$.	
\par
One should pause and appreciate the generality of Eq.~\eqref{div}.
Recall, in particular, that we have not specified any particular model.
Any theory~\footnote{We have displayed the calculation for bosons, but the same results
can be easily shown to hold for fermions (up to global signs).}
whose Hessian of the classical action is of the form~\eqref{hessian}
has one-loop divergences given by Eq.~\eqref{div}.
Finding the one-loop divergences of a certain theory thus boils down to the knowledge
of the potential term $\hat P$ and the connections that define the curvatures in
Eqs.~\eqref{curv1} and \eqref{curv2}.
One should also note that the spacetime connection in Eq.~\eqref{hessian}
has been assumed torsionless (an apparent contradiction to our purposes)
and metric-compatible.
In general, scalars formed by the torsion and non-metricity tensors would also
show up in Eq.~\eqref{div} (see {\em e.g.}, Ref.~\cite{Obukhov:1983mm}).
However, as we have seen in Section~\ref{TG}, in teleparallel gravity one can phrase torsion
effects either in terms of the Levi-Civita Ricci scalar [see Eq.~\eqref{QR}]
or in terms of the gauge field strength for the translation group.
In the former scenario, which takes place when the full-coupling prescription (gauge + local Lorentz)
is adopted, the result is equivalent to general relativity, albeit expressed in terms of different
quantities.
This is rather expected because the full-coupling prescription is equivalent to the coupling
obtained from the equivalence principle in general relativity.
On the other hand, when one only imposes the gauge-coupling prescription, the UV divergences
in the latter case are captured by the last term in Eq.~\eqref{div}, which is the typical result
of gauge theories.
This is indeed the crucial point that can render quantum fields in teleparallel gravity renormalizable
at one-loop order by introducing a slight modification in the teleparallel gravity action.
\section{One-loop divergences in teleparallel gravity}
\setcounter{equation}{0}
\label{section TEGR 1-loop corrections}
This section concerns the application of the formalism reviewed in Section~\ref{SD}
to teleparallel gravity.
As we have seen in Section~\ref{TG}, there are two distinct connections in teleparallel gravity,
namely the spin connection $\overset{\bullet}{\omega}\,\!^a_{\ \mu\nu}$, which accounts for the
local Lorentz invariance, and the gauge field $B^a_{\ \mu}$, which accounts for the
gauge invariance under bundle translations.
The coupling to matter then follows a two-step prescription, which results in the standard
equations of motion for matter fields in curved spacetime (except that spacetime is no longer
curved but twisted).
\par
Needless to say, the coupling to matter is a crucial physical aspect of the theory
and turns out to have important consequences for the quantization of both sectors.
It is indeed the decisive factor for the renormalizability of the theory.
In the following, we shall analyse two possibilities for the matter coupling and compute
the one-loop divergences in each case.
We first consider the full-coupling prescription, obtained from both gauge and Lorentz symmetries.
The Schwinger-DeWitt expansion in this case thus comprises geometrical invariants constructed
with both connections.
For the second case, we take into account only the gauge invariance in order to define the coupling
prescription, following the standard procedure of gauge theories.
\subsection{Full-coupling prescription}
For simplicity, we shall consider a scalar field coupled to gravity.
This is enough for our purposes and considerations for other types of fields follow directly from
Eqs.~\eqref{HAMIDEW0}-\eqref{HAMIDEW2}.
The action for a massive scalar field minimally coupled to gravity is obtained from the full
gravitational coupling prescription~\eqref{fullc} as
\begin{equation}
	S_\phi
	=
	-\frac1 2
	\int d^{4}x\,
	h\,
	\left(
    g^{\mu\nu}\, \overset{\bullet}{\nabla}_{\mu}\phi \,\overset{\bullet}{\nabla}_{\nu}\phi
    + m^2 \phi^2
    \right)
	\ ,
\end{equation} 
where $h$ is again the determinant~\eqref{deth}.
In this case, the operator of interest reads
\begin{equation}
	F(\overset{\bullet}{\nabla})
	=
	\overset{\bullet}{\nabla}_{\mu}\,\overset{\bullet}{\nabla}\,\!^{\mu}
	-m^2
	\ .
	\label{scalar_diffOp}
\end{equation} 
Let us recall that the connection in $\overset{\bullet}{\nabla}\,\!_{\mu}$ only shows up in the
combination
$\overset{\circ}{\omega}\,\!^{ab}_{\ \ \mu} = \overset{\bullet}{\omega}\,\!^{ab}_{\ \ \mu} - \overset{\bullet}{K}\,\!^{ab}_{\ \ \mu}$,
thus reproducing the same effects of the Levi-Civita connection in general relativity.
In particular, since the one-loop divergences depend solely upon geometrical invariants
built with the underlying connection, one would expect the same sort of divergent structure
when matter fields are coupled to teleparallel gravity.
In fact, by recalling the relation~\eqref{QR}, the commutator of covariant derivatives in this case reads
\begin{equation}
	\left[ \overset{\bullet}{\nabla}\,\!_{\mu}, \overset{\bullet}{\nabla}\,\!_{\nu}\right] V^{\alpha}
	=
	-\overset{\,\, \bullet}{Q}\,\!^{\alpha}_{\ \beta\mu\nu}\,V^{\beta}
	=
	\overset{\,\, \circ}{R}\,\!^{\alpha}_{\ \beta\mu\nu}\,V^{\beta}
	\ ,
	\label{com}
\end{equation}
for any spacetime vector $V^\alpha$.
We remark that the Schwinger-DeWitt coefficients shown in Eqs.~\eqref{HAMIDEW0}-\eqref{HAMIDEW2}
are only valid for a torsionless connection.
In the presence of torsion, the situation becomes more involved with many new terms.
Nonetheless, since teleparallel geometry satisfies Eq.~\eqref{QR}, one can use the result~\eqref{com}
and replace all instances of the Riemann tensor $\overset{\circ}{R}\,\!^{\alpha}_{\ \beta\mu\nu}$
with $-\overset{\bullet}{Q}\,\!^{\rho}_{\ \lambda\mu\nu}$. 
Moreover, there is no potential or gauge field strength in this case, thus the coefficients~\eqref{HAMIDEW0}-\eqref{HAMIDEW2}
become 
\begin{equation}
	\begin{split}
		a_{0}(x,x)
		&=
		1 
		\\
		a_{1}(x,x)
		&=
		{-\frac{\overset{\,\, \bullet}{Q}}{6}}
		\\
		a_{2}(x,x)
		&=
		\frac{1}{180}\left(
		\overset{\bullet}{Q}\,\!_{\alpha\beta\mu\nu} \,\overset{\bullet}{Q}\,\!^{\alpha\beta\mu\nu}  
		- \overset{\bullet}{Q}\,\!_{\mu\nu}\,\overset{\bullet}{Q}\,\!^{\mu\nu}
		+\frac{5}{2} \,\overset{\,\, \bullet}{Q}\,\!^{2} 
		- 6\,  \overset{\bullet}{\nabla}\,\!_{\mu}\overset{\bullet}{\nabla}\,\!^{\mu}  \overset{\bullet}{Q}
		\right)
		\ .
		\label{HAMIDEW_coeff}
	\end{split}
\end{equation}
Note that the tensor $\overset{\bullet}{Q}\,\!^{\rho}_{\ \lambda\mu\nu}$ is quadratic in the torsion tensor,
thus $a_2$ contains terms of fourth order in the torsion, whereas the bare Lagrangian is only quadratic.
This is indeed reminiscent of the non-renormalizability of the matter sector in general relativity.
\subsection{Gauge coupling only}
\label{gaugeonly}
One could argue that the most natural choice for the gravitational coupling is obtained by following
the usual gauge prescription and replacing partial derivatives by gauge covariant ones, that is
\begin{equation}
	\partial_{\mu} \phi \to h_{\mu} \phi
	=
	\partial_\mu \phi
	+ {\sqrt{16\,\pi \, G_{\rm N} } } B^a_{\ \mu} \,\partial_a \phi
	\ .
	\label{gauge_grav_coupling}
\end{equation}
In this case, instead of Eq.~\eqref{scalar_diffOp}, one obtains the relevant operator
\begin{equation}
    F(h)
    =
    h_\mu \,h^\mu - m^2
    \ .
\end{equation}
It is important to stress that $h_\mu$ does not contain a spacetime connection,
but it rather involves only the genuine gauge field $B^a_{\ \mu}$.
The form of the one-loop divergences correspondingly change dramatically.
From Eq.~\eqref{HAMIDEW0}-\eqref{HAMIDEW2}, we now find
\begin{equation}
	\begin{split}
		a_{0}(x,x)
		&=1
		\\
		a_{1}(x,x)
		&=0
		\\
		a_{2}(x,x)
		&=
		\frac{1}{12}\, \hat{\mathcal{R}}_{\mu\nu} \,\hat{\mathcal{R}}^{\mu\nu}
		\ ,
		\label{newHAMIDEW_coeff}
	\end{split}
\end{equation}
where $\hat{\mathcal R}_{\mu\nu}$ is the translational field strength, namely the torsion tensor
$\hat{\mathcal R}_{\mu\nu}=\overset{\bullet}{T}\,\!^{a}_{\ \mu\nu}\,\partial_{a}$.
\par
Note that not all possible contractions of the torsion tensor show up in $a_2$,
but only the one corresponding to the first term in Eq.~\eqref{TEGR}.
The early-time asymptotic expansion of the heat kernel indeed comprises only commutators
of covariant derivatives, thus reproducing only the standard result of Yang-Mills theories.
Moreover, the divergent part in the present case is quadratic in the torsion tensor rather than quartic
as before, which might mislead one to think that the theory could be renormalizable.
However, the bare action contains only Newton's constant, so that the one-loop divergences could only
be absorved by the bare Newton constant if they appeared in the exactly same combination
as in Eq.~\eqref{TEGR}.
This clearly suggests an easy-fix should one depart from the formulation that is equivalent
to general relativity and consider general teleparallel theories in which the different terms
in the bare action are accompanied by different coupling constants.
For instance, we could consider
\begin{equation}
    \overset{\,\, \bullet}{\mathcal{L}}
	=
    {-\frac{h}{16\,\pi\,G_{\rm N}} }
	\left(
	    c_1 \,\overset{\bullet}{T}\,\!^{\rho}_{\ \mu\nu} \,\overset{\bullet}{T}\,\!_{\rho}^{\ \mu\nu}
	    + c_2\, \overset{\bullet}{T}\,\!^{\rho}_{\ \mu\nu} \,\overset{\bullet}{T}\,\!^{\nu\mu}_{\ \ \rho}
	    + c_3 \,\overset{\bullet}{T}\,\!^{\rho}_{\ \mu\rho}\,\overset{\bullet}{T}\,\!^{\nu\mu}_{\ \ \nu}
	\right)
	\ ,
	\label{genTG}
\end{equation}
where $c_i$ are dimensionless constants.
In this case, one can renormalize all divergences at one-loop without having to include higher-order terms, 
thus suggesting that the theory can be renormalizable, while precluding the existence of ghosts.
The proof of renormalizability, however, requires the extension of the above result to all loop orders.
\par
One can still preserve the equivalence with general relativity at the classical level.
Since divergences appear at one-loop order, the necessary counter-terms must be proportional to the 
Planck length squared, as shown in Eq.~\eqref{gammaLp}.
We can therefore have 
\begin{equation}
c_i
\simeq
b_i+\ell_{\rm p}^2\,d_i
\ ,
\label{cdLp}
\end{equation}
with $b_1=1/4$, $b_2=1/2$ and $b_3=-1$ so as to recover Eq.~\eqref{TEGR} for $\ell_{\rm p}\to 0$
(equivalent to $\hbar\to 0$).
Note that the coefficients $d_i$ must have dimensions of $\ell_{\rm p}^{-2}$.
This is the dimension of the field strength $\hat{\mathcal{R}}_{\mu\nu}$, which determines the 
non-trivial coefficient $a_2$ in Eq.~\eqref{newHAMIDEW_coeff}.
Without gravity, one has vanishing torsion $\hat{\mathcal{R}}_{\mu\nu}=0$ and no divergence
appears.~\footnote{We recall the comment in footnote~\ref{f:1} about the use of a background geometry
(or torsion) to define the effective actions.}
On the other hand, if $\hat{\mathcal{R}}_{\mu\nu}\not=0$, there must be a matter source
to accelerate (test) particles and the coefficients $d_i$ will then be related to the typical size
of those sources by the proper field equations.
This is indeed qualitatively similar to quantum general relativity, where one-loop divergences also 
depend on the presence of matter sources and vanish in flat space with $R_{\mu\nu\alpha\beta}=0$.
\par
General relativity contains only Newton's coupling constant $G_{\rm N}$, both for matter-gravity
interactions and gravity self-interactions.
On the other hand, the most general teleparallel Lagrangian~\eqref{genTG} contains three couplings
$G_i=G_{\rm N}/c_i$ for the gravitational self-interaction.
With the condition~\eqref{cdLp}, the two theories coincide at the classical level but separate
at the quantum level.
One can further envisage a modification in the quantum matter-gravity sector as well,~\footnote{For
some recent proposals, see {\em e.g.},~Refs.~\cite{Feng:2019dwu,Casadio:2019cux}.}
depending on which of the $G_i$ appear in the interaction with matter fields.
\par
In particular, we stress that the only difference between the Lagrangian~\eqref{genTG} and the model of the
previous section regards the presence of local Lorentz invariance and the number of free parameters.
Comparing both results points to the local Lorentz invariance and the lack of additional coupling constants
as the culprits for the non-renormalizability of general relativity.
This observation is only possible because of the separation of the Levi-Civita connection into an inertial part
and a pure gravitational one, ultimately giving rise to two distinct possibilities for the coupling prescription.
\par
There are important differences between these coupling prescriptions that one should keep in mind.
For one, the gauge-coupling prescription cannot be abandoned as it is the true responsible for coupling
the gravitational field $B^a_{\ \mu}$ to matter, thus introducing dynamical degrees of freedom.
Local Lorentz symmetry, on the other hand, plays no dynamical role, and is required only to enforce
frame-independence. 
Moreover, the metric remains invariant regardless of the chosen frame since
$g_{\mu\nu} = \eta_{ab}\, h^a_{\ \mu}\, h^b_{\ \nu}$ is invariant under local Lorentz transformations.
Hence local measurements of the gravitational field cannot detect violations of the local Lorentz symmetry.
Such violations would only become detectable by observations in the matter sector.
\par
It is also important to distinguish between local and global Lorentz invariance.
By giving up on the full-coupling prescription, violations of the former, but not of the latter, are allowed.
In addition, local Lorentz invariance is not an exact symmetry of the the teleparallel action~\eqref{TEGR}
in that it only holds up to a boundary term, whereas the global Lorentz symmetry is exact.
Although boundary terms do not affect the equations of motion, their importance cannot be overlooked
in the quantum (or even semiclassical) regime where non-stationary configurations also contribute
to the path integral.
\par
The above discussion suggests a different route for the quantization of quantum fields coupled to gravity.
One should adopt Eq.~\eqref{genTG} as the action for the gravitational sector.
The gauge-coupling prescription then induces the interaction of gravity with the other sectors.
At this point, one first quantizes all matter fields and only then imposes local Lorentz symmetry
among the couplings.
This guarantees that the theory is renormalizable at one-loop order while retaining local Lorentz
invariance.
\section{Conclusions}
\setcounter{equation}{0}
\label{s:conc}
In this paper, we have studied the one-loop divergences produced by quantum fields coupled
to a classical background described by teleparallel gravity. 
The presence of both local Lorentz symmetry and translational gauge invariance requires
a two-step coupling prescription.
We considered both coupling prescriptions in isolation in order to track down the origin of
the non-renormalizability of quantum fields in general relativity.
We have shown that the theory of quantum fields cannot be made one-loop renormalizable,
in any circumstance, when the background is described by the teleparallel equivalent
to general relativity.
However, the formulation of teleparallel gravity clearly shows where things go wrong.
An one-loop renormalizable theory is possible if one replaces the coefficients in Eq.~\eqref{TEGR}
with the free parameters in Eq.~\eqref{genTG} and quantizes the matter sector before imposing
local Lorentz invariance.
We note that the gauge invariance is responsible for giving rise to dynamical fields,
whereas the local Lorentz symmetry has no dynamical content whatsoever.
Therefore, the gauge-coupling prescription, followed by quantization and only then followed by
the local-Lorentz-coupling prescription is rather natural.
\par
Although the action~\eqref{genTG} is no longer equivalent to general relativity,
the two theories can be kept equivalent at the classical level and
the quantization of matter fields in such a background does not require higher-derivative terms.
Contrary to Stelle's fourth-derivative gravity~\cite{Stelle:1976gc}, for example, the renormalization
at one-loop does not introduce instabilities or violations of unitarity.
We stress, however, that we have only considered divergences at one-loop order,
and the full renormalizability of the theory is left for future works.
We have not quantized gravity (here represented by $B^a_{\ \mu}$) either.
Nonetheless, we expect that the quantization of $B^a_{\ \mu}$ only modifies
the coefficients of the divergences as usual.
The details of such a calculation shall however be presented elsewhere.
\section*{Acknowledgments}
R.C.~and I.K.~are partially supported by the INFN grant FLAG.
The work of R.C.~has also been carried out in the framework of activities of the National Group
of Mathematical Physics (GNFM, INdAM). 

\end{document}